\documentclass[jgrga]{agutex2015}

\usepackage{color,soul}
\usepackage{graphicx}
\usepackage{amsmath}
\usepackage{rotating}
\newcommand{\blue}{\textcolor{black}}
\setkeys{Gin}{draft=false}

\authorrunninghead{ SHEN ET AL.}

\titlerunninghead{Ion Heating By BBELF WAVES}

\authoraddr{Corresponding author: Yangyang Shen, Department of Physics and Astronomy, University of Calgary, Calgary, Alberta, Canada. (yangyang.shen@ucalgary.ca)}

\begin{document}


\title{On O$^{+}$ ion heating by BBELF waves at low altitude: Test particle simulations }




\authors{Yangyang Shen\altaffilmark{1, 2},
David J. Knudsen\altaffilmark{1}}

\altaffiltext{1}{Department of Physics and Astronomy, University of Calgary, Calgary, Alberta, Canada.}
\altaffiltext{2}{Now at Department of Earth, Planetary, and Space Sciences, University of California, Los Angeles, California, USA}



\begin{abstract}

We investigate mechanisms of wave-particle heating of ionospheric O$^{+}$ ions resulting from broadband extremely low frequency (BBELF) waves using numerical test particle simulations that take into account ion-neutral collisions, in order to explain observations from the Enhanced Polar Outflow Probe (e-POP) satellite at low altitudes ($\sim$400 km) \citep{shen2018}. We argue that in order to reproduce ion temperatures observed at e-POP altitudes, the most effective ion heating mechanism is through cyclotron acceleration by short-scale electrostatic ion cyclotron (EIC) waves with perpendicular wavelengths $\lambda_{\perp}$$\leq$200 m. The interplay between finite perpendicular wavelengths, wave amplitudes, and ion-neutral collision frequencies collectively determine the ionospheric ion heating limit, which begins to decrease sharply with decreasing altitude below approximately 500 km, where the ratio $\frac{\nu_c}{f_{ci}}$ becomes larger than 10$^{-3}$, $\nu_c$ and $f_{ci}$ denoting the O$^+$--O collision frequency and ion cyclotron frequency. We derive, both numerically and analytically, the ion gyroradius limit from heating by an EIC wave at half the cyclotron frequency. The limit is 0.28$\lambda_{\perp}$. The ion gyroradius limit from an EIC wave can be surpassed either through adding waves with different $\lambda_{\perp}$, or through stochastic ``breakout'', meaning ions diffuse in energy beyond the gyroradius limit due to stochastic heating from large-amplitude waves. Our two-dimensional simulations indicate that small-scale (\textless1 km) Alfv{\'e}n waves cannot account for the observed ion heating through trapping or stochastic heating.

\end{abstract}


\begin{article}

\section{Introduction}

 The importance of broadband extremely low-frequency (BBELF) waves for transverse ionospheric ion heating and ion outflow in the high-latitude auroral region has been well established \citep{kintner1996, bonnell1996, knudsen1998,lynch2002}. Based on observations from sounding rockets and satellites, a number of wave modes may exist in the broadband emission from several Hz up to several kHz, including Alfv{\'e}n waves below the ion cyclotron frequency \citep{stasiewicz2000a, chaston2004}, ion acoustic waves \citep{wahlund1998}, electrostatic waves near the ion cyclotron frequency (EIC waves) \citep{kintner1996,bonnell1996}, and electromagnetic ion cyclotron (EMIC) waves \citep{erlandson1994}. 
 
 Although it is agreed upon that perpendicular electric fields, either left-hand circularly polarized or linearly polarized, are essential for ion acceleration \citep{lysak1980,chang1986,chaston2004}, the ion heating mechanism, dominant wave modes, and perpendicular wavelengths that are responsible for ion heating are still not well known \citep{paschmann2002}. Until recently, BBELF heating was thought to occur well above the ionosphere where collisional effects are negligible \citep[e.g.][]{wu1999,zeng2006}. \citet{shen2018} reported multiple examples of BBELF-induced heating at altitudes as low as 350 km and at spatial scales as narrow as 2 km. These observations call for a more comprehensive and realistic ion heating model in the BBELF wave environment, including the effects of both ionospheric ion-neutral collisions and finite perpendicular wavelengths. There has been no such model in the literature yet.

Previous studies have suggested different mechanisms of ion heating from auroral plasma waves. Ions can be resonantly accelerated by lower-hybrid waves when the perpendicular wave phase velocity is comparable with the ion velocity, leading to a heated tail \citep{lynch1996,tsurutani1997}. On the other hand, particles in the bulk of the ion population can be accelerated simultaneously by long-wavelength (much larger than the ion gyroradius $\rho_{i}$) ion cyclotron waves through quasi-linear cyclotron resonance \citep{chang1986}. \citet{ball1991} used test particle simulations to study ion heating by BBELF waves and found that the heating rate increases as the wave frequency approaches the ion cyclotron frequency.

A similar, but nonlinear, acceleration mechanism is the coherent ion trapping by electrostatic ion cyclotron waves \citep{lysak1980,ram1998}. \citet{lysak1980} found that, in a single electrostatic ion cyclotron wave, ions will be trapped in an effective potential well. The energy, or equivalently the gyroradius, of the ion is limited by the first zeros of the Bessel function of the first kind, governed by the parameter $k_{\perp}$$\rho_{i}$, where $k_{\perp}$ is the perpendicular wavenumber. \citet{lysak1980} also showed that the presence of additional waves with different wavenumbers can detrap the ion, and the gyroradius barrier can be surpassed. \citet{lysak1986} extended this theory to harmonics of EIC waves, in which case the upper limits of the ion gyroradius are specified by the first zeros of $J_{n}(k_{\perp}\rho_{i})$, where $n$ is the harmonic integer order. In this paper, we extend these previous results of gyroradius limits to EIC waves with the frequency of $\omega_c \over 2$, which has not been reported yet. We shall provide consistent analytical and numerical calculations for the gyroradius limits.

Another important type of acceleration mechanism is stochastic ion heating, in which ion energy increases due to random kicks in phase space, increasing temperature. Stochastic ion heating has been studied in detail for EIC waves and lower-hybrid waves \citep{karney1978, papa1980, lysak1986}. The onset condition is: $\frac{E_{\perp}}{B_0} \simeq {\frac{1}{4}}{\left( \frac{\Omega_{i}}{\omega} \right)}^{\frac{1}{3}} {\frac{\omega}{k_{\perp}}}$, where $E_\perp$ is the amplitude of perpendicular wave electric field, $B_0$ the magnitude of magnetic field, $\Omega_{i}$ the ion cyclotron frequency, and $\omega$ the wave angular frequency. 

Stochastic ion heating due to large-amplitude Alfv{\'e}n waves has also been investigated by several subsequent studies \citep{mcchesney1987,bailey1995,stasiewicz2000b,chen2001,johnson2001,chaston2004}. \citet{chaston2004} found that, for transverse wave amplitude $E_{\perp}$ satisfying ${E_{\perp} \over {B_0}} \textless {\Omega_{i} \over k_{\perp}}$, ion motion in the wave field is coherent and the ion may become trapped by Alfv{\'e}n waves in a manner similar to that proposed for EIC waves. The ion may be accelerated up to a gyrodiameter roughly equivalent to the perpendicular wavelength of the wave. When ${E_{\perp} \over {B_0}} \textgreater {\Omega_{i} \over k_{\perp}}$, the ion can be stochastically accelerated, and the gyroradius limit can be surpassed. In addition, in the presence of a large spatial gradient of electric fields, significant ion acceleration and orbit chaotization may take place \citep{cole1976,stasiewicz2000b}. However, amplitudes of Alfv{\'e}n or EIC waves observed in the low-altitude ionosphere are generally insufficient to initiate stochastic ion heating, according to the onset condition of the stochastic behavior \citep{karney1978, bonnell1996,chaston2004}. 

Ion-neutral collisions, due to large DC electric fields in the ionosphere, may generate a significant amount of O$^+$ ion heating and sometimes lead to anisotropic and toroidal ion distributions \citep{maurice1979,loranc1994,wilson1994}. However, collisions can also put a limit on the ion heating in the ionosphere by restricting the achievable speed relative to neutral gases \citep{schunk_nagy2009}. Few models of ionospheric ion heating by plasma waves take into account both the effects of finite perpendicular wavelengths and ion-neutral collisions. Using a multispecies particle simulation, \citet{providakes1990} showed that collisional current-driven electrostatic ion cyclotron waves may be unstable in the bottomside ionosphere ($<$300 km) and that they are able to generate transverse bulk acceleration of the heavy ions to energies of more than a few keV in this region. However, the required critical field-aligned current for the cyclotron instabilities is at least 50 $\mu$A/m$^2$, which is seen rarely at best, and is significantly higher than values reported by \citet{shen2018}. \citet{burleigh2018} used an anisotropic fluid model to study ion upflows in the ionosphere from drivers of both quasilinear wave-ion heating and collisions. But this fluid model simplifies wave-ion heating processes and does not incorporate finite wavelength effects.  

In this paper, we shall explore, using test particle simulations, whether and how cyclotron and stochastic ion heating from Alfv{\'e}n (planar or non-planar) or EIC waves with different perpendicular wavelengths, along with ion-neutral collisions, might contribute to ionospheric O$^{+}$ ion heating to observed levels, as exemplified by those found from the Enhanced Polar Outflow Probe (e-POP) at 410 km altitude \citep{shen2018}. The test particle simulations we use are admittedly simple but allow us to understand the microphysical (kinetic) processes in some detail. In the following, we shall first examine two different ion heating mechanisms---stochastic acceleration and coherent trapping---and then investigate the role of BBELF waves along with ion-neutral collisions in explaining O$^{+}$ ion heating under the constraints of e-POP observations. 


\section{Stochastic ion acceleration by a monochromatic electrostatic wave}

We assume a coherent monochromatic electrostatic wave in a uniform background magnetic field $\vec{B} = B_0 \hat{z}$:
\begin{equation}
\label{eq:Efield}
\vec{E} = E_0 \text{cos}(kx-{\omega}t) \hat{x}
\end{equation}
where $B_0$ is the magnitude of magnetic field, $E_0$ the amplitude of the electrostatic wave, $k$ the wavenumber, and $\omega$ the wave angular frequency.

We use the Hamiltonian method to describe the charged particle's motion perpendicular to the background magnetic field. The Hamiltonian reads:
\begin{equation}
\label{eq:H}
H = T_{kinetic} + V_{potential} = \frac{1}{2m}[(P_{y} + q{B_0}x)^2 + {P_{x}}^2] - \left({q{E_0} \over k}\right) \text{sin}(kx-{\omega}t)
\end{equation}
where $P_{x}=mv_x$ and $P_{y}=mv_y$ are the ion momentum terms. Following previous similar studies \citep{karney1978,bailey1995}, we can normalize the physical quantities as $t'=\Omega_{i}t$, $x'=kx$, ${P_{y}}'=P_{y}{k \over m\Omega_{i}}$, ${P_{x}}'=P_{x}{k \over m\Omega_{i}}$, ${\alpha}={\frac{E_0}{B_0} \frac{k} {\Omega_i}}$, and $\nu={\omega \over \Omega_{i}}$, where the primed variables indicate normalized quantities. $\Omega_{i}={q{B_0} \over m}$ is the ion cyclotron frequency, $\alpha$ the normalized wave potential, and $\nu$ the normalized wave frequency. The Hamiltonian takes this new form:
\begin{equation}
\label{eq:nH}
H = \frac{1}{2}(P_{y} + x)^2 + \frac{1}{2}{P_{x}}^2 - \alpha \text{sin}(x-{\nu}t)
\end{equation}
where we have dropped the prime symbol. The equations of motion for the particle can be found as:
\begin{equation}
\label{eq:x}
\dot{x} = {\partial{H} \over \partial{P_{x}}} = P_{x}
\end{equation}
\begin{equation}
\label{eq:Px}
\dot{P_{x}} = -{\partial{H} \over \partial{x}} = -P_{y} - x + \alpha \text{cos}(x-{\nu}t)
\end{equation}
\begin{equation}
\label{eq:y}
\dot{y} = {\partial{H} \over \partial{P_{y}}} = P_{y} + x
\end{equation}
\begin{equation}
\label{eq:Py}
\dot{P_{y}} = -{\partial{H} \over \partial{y}} = 0
\end{equation}
where we can identify that $P_{y}$ is a constant of motion; therefore, the dynamics in terms of $x$ and $P_{x}$ are exclusively determined by each other. The symmetric equations can be conveniently integrated numerically using the 4th-order symplectic integrator \citep{forest1990}. Such an integrator is phase-space area conserving and more suitable for long-term integration. It also takes less computing time than the classic Runge-Kutta scheme. The orbit of a particle in phase space is often represented by a Poincar{\'e} surface of section plot, constructed by marking the particle's trajectory when it passes a constant plane in phase space, e.g., when the wave phase equals $2\pi$ in our case. We use 30 test particles that are uniformly distributed in velocity space to explore their trajectories under the influence of the wave electric field. The accuracy of the 4th order symplectic integrator has been tested on a simple harmonic oscillator so that the relative trajectory error is less than 0.01\% for 10 million integration steps. A stepsize of 0.01 (equivalent to $\frac{1}{100}{\Omega_{i}}^{-1}$) is chosen as we numerically calculate the particle's motion over two million steps and construct the Poincar{\'e} plots.

In the following, we define the off-resonance case when $\nu=0.1$ and the on-resonance case when $\nu=1$. Linearly polarized waves having a frequency much less than the ion cyclotron frequency ($\nu=0.1$) may represent Alfv{\'e}n waves in the sheet-like current structures in the auroral region. In the case described here, when the perpendicular wave electric field and wave vector are approximately parallel, Alfv{\'e}n waves are in the shear mode. Throughout the paper, the term ``Alfv{\'e}n wave'' therefore refers to a linearly polarized wave with a frequency much lower than the ion cyclotron frequency. We ignore the magnetic perturbation, which is justifiable because magnetic perturbations at low altitudes ($\sim$400 km) are insignificant compared with the background magnetic field. The on-resonance case corresponds to an EIC wave. Both cases are investigated for different wave potentials or amplitudes $\alpha$, in order to understand the ion stochastic motion in the Alfv{\'e}nic and ion cyclotron regimes. 

\begin{figure}[h]
\includegraphics[width=9cm,scale=0.6,trim={10cm 1cm -4cm 3cm}]{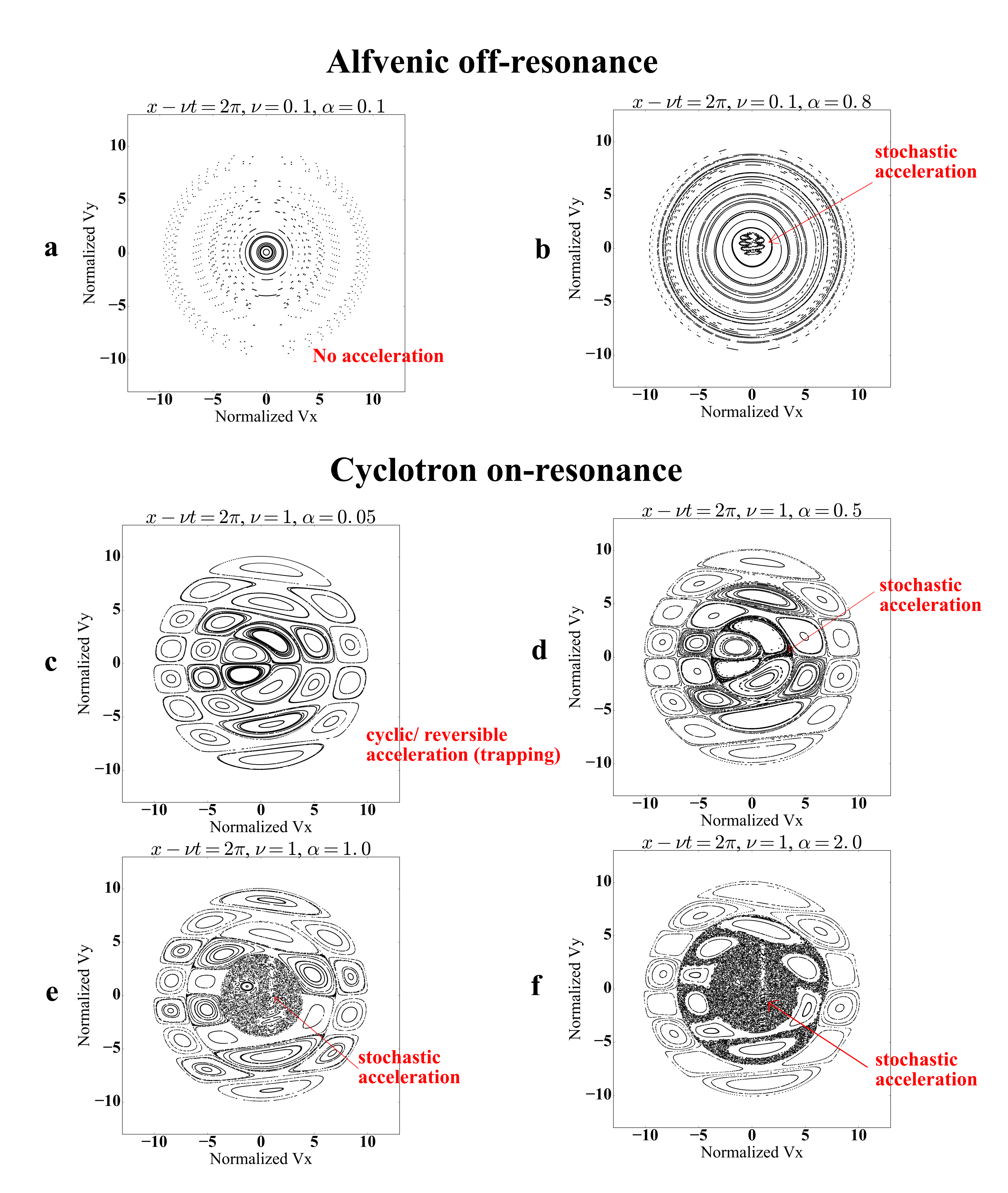}
\caption{Poincar{\'e} surface of section plots for both Alfv{\'e}nic off-resonance (Figure 1a, 1b) and electrostatic ion cyclotron (EIC) wave on-resonance (Figure 1c, 1d, 1e, 1f) cases for different values of wave amplitude $\alpha$. For Alfv{\'e}n waves, there is no net ion energy increase when the wave amplitude is small ($\alpha=0.1$) (Figure 1a). However, when $\alpha$ approaches 0.8, the ions can undergo stochastic acceleration as shown in Figure 1b. For the on-resonance cases, when $\alpha$ gradually increases from 0.05 to 2.0, we observe that stochastic motion emerges from the inner circle and expands into outer circles. The largest circle is specified by $J_{1}(k\rho_{i})=0$.}
\label{stochastic}
\end{figure}

Figure~\ref{stochastic} shows Poincar{\'e} surface of section plots for both off- (Figure~\ref{stochastic}a, \ref{stochastic}b) and on-resonance (Figure~\ref{stochastic}c, \ref{stochastic}d, \ref{stochastic}e, \ref{stochastic}f) cases when the wave amplitude $\alpha$ is set to different values. In Alfv{\'e}n waves, there is no net ion energy increase when the wave amplitude is small ($\alpha=0.1$). However, when $\alpha$ approaches 0.8, or equivalently, ${{\frac{E_0}{B_0}}{\frac{k}{\Omega_{i}}}}\simeq0.8$, the ion can undergo stochastic acceleration. Figure~\ref{stochastic}b shows such a phonomenon, where ions start to randomly occupy the velocity space within a specific circle, which represents the energy limit that ions can gain. The stochastic onset condition for Alfv{\'e}n wave ion heating is consistent with \citet{chaston2004}. Within the stochastic region, ions have access to all the available phase space region, or equivalently, ions can be accelerated to any energy level accessible. 

An ion trajectory that follows a circle in phase space means that its energy does not increase. However, for EIC waves, the ion's energy increases and returns back to its original level cyclically (Figure~\ref{stochastic}c) if the ion stays in the system for a sufficiently long time. In this case, the ion is trapped within wave potential wells. The energy limit is determined by harmonic potential structures of the EIC wave, which can be represented by expansion of the wave potential in terms of the Bessel function ${J_n}(k{\rho_{i}})$ of the first kind \citep{lysak1980,gibelli2010}. Similar to Figure~\ref{stochastic}b, when $\alpha$ gradually increases from 0.05 to 2.0, we observe that stochastic ion motion emerges from the inner circle and expands into outer circles. The largest circle is specified by $J_{1}(k\rho_{i})=0$ \citep{lysak1980}. The stochastic onset condition for EIC waves was numerically determined by \citet{karney1978} as $\frac{E_{\perp}}{B_0} \simeq 0.25{\left( \frac{\Omega_{i}}{\omega} \right)}^{\frac{1}{3}} {\frac{\omega}{k_{\perp}}}$. The numerical constant we found instead is 0.4, which is similar to 0.25 found earlier using different scenarios and parameters. The formula derived from \citet{karney1978} concerns electrostatic waves with frequencies much larger than the ion cyclotron frequency.  



\section{Coherent ion cyclotron acceleration by EIC waves}

To understand ion cyclotron acceleration from EIC waves with different frequencies and perpendicular wavelengths, we perform another numerical simulation using the classic Runge-Kutta integrator to resolve ions dynamics. This is because we aim later to investigate how different frequency and wavenumber spectra may affect ion bulk heating, as applicable to the BBELF-heating scenario. The symplectic integrator has no advantage in this situation since the problem is not symmetric. We shall report two phenomena of ion acceleration in the single wave case, i.e., the ion gyroradius limit and the stochastic ``breakout'', where at a certain critical point an ion's energy can diffuse beyond the gyroradius limit through stochastic acceleration. 

The ion dynamics perpendicular to the background magnetic field are determined by the Lorentz equation: 
\begin{equation}
\label{eq:lorentz}
{m\frac{\partial{\vec{v}}}{\partial{t}}} = {q(\vec{E}(x,t)+\vec{v}\times\vec{B})}
\end{equation}
where $\vec{B}$ only has a $\hat{z}$ component and $\vec{E}(x,t)$ is the wave electric field in the $\hat{x}$ direction. For a single wave case, $\vec{E}(x,t)={E_{0}}\text{cos}({\frac{2\pi}{\lambda_{\perp}}}x-{\omega}t)\hat{x}$. The equations of motion in the perpendicular directions are:
\begin{equation}
\label{eq:Vx}
\dot{v_{x}} = {\frac{q}{m}}\vec{E}(x,t)+{\Omega_{i}}{v_{y}}
\end{equation}
\begin{equation}
\label{eq:Vy}
\dot{v_{y}} = -{\Omega_{i}}{v_{x}}
\end{equation}
 where $\Omega_{i}$ is the ion gyrofrequency. The equation can be numerically integrated using the 4th-order Runge-Kutta integrator. The accuracy of the integrator has been tested by comparing numerical results with the analytical solution of the $E{\times}B$ drift. A stepsize of $0.01{T_{gyro}}$ is chosen to limit the relative $E{\times}B$ trajectory error to be within $10^{-6}$ over 10 million integration steps. 
 
\begin{figure}[h]
\includegraphics[scale=0.6,width=8cm,trim={1cm 1cm 0 2cm}]{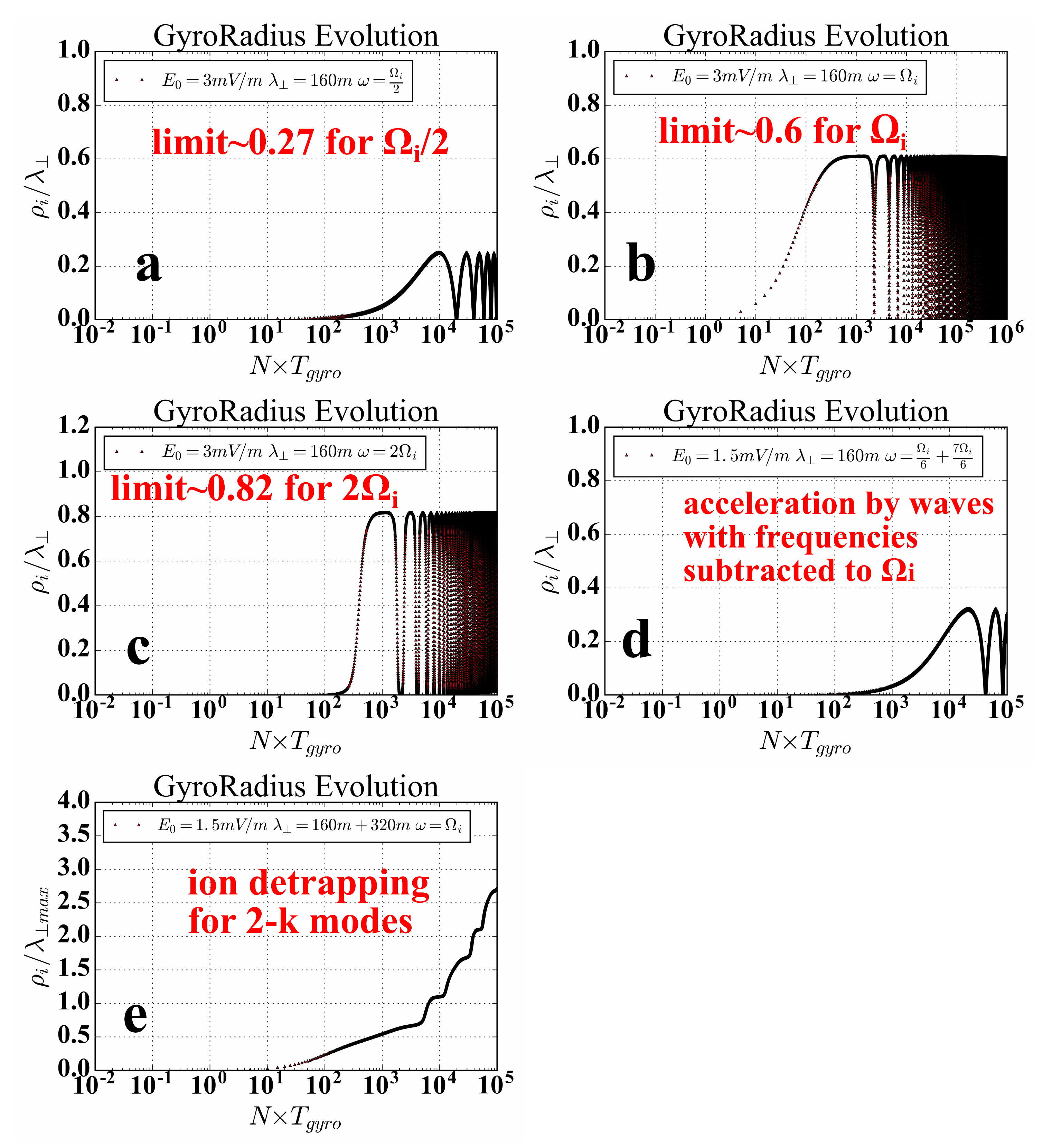}
\caption{Time evolution of the ratio $\frac{\rho_{i}}{\lambda_{\perp}}$ for EIC waves with $\omega={\frac{\Omega_i}{2}}$, ${\Omega_i}$, and $2{\Omega_i}$, and for the cases with two different wavelengths and two frequencies. The ratio limits are approximately 0.27, 0.6, and 0.83 in Figure 2a, 2b and 2c. Figure 2d shows a moderate ion radius limit ($\sim$0.3) for waves with two frequencies whose difference is the cyclotron frequency. In Figure 2e, the ion gyroradius exceeds the limits set by both perpendicular wavelengths. }
\label{gyrolimit}
\end{figure}

In the presence of a single cyclotron wave, an ion is accelerated over the entire gyroorbit; the ion's velocity has a component in the same direction of the wave electric field. Figure~\ref{gyrolimit}a, \ref{gyrolimit}b, and \ref{gyrolimit}c show the gyroradius evolution for $\omega={\frac{\Omega_i}{2}}$, ${\Omega_i}$, and $2{\Omega_i}$. The $\rho_{i} \over\lambda_{\perp}$ limits are approximately 0.27, 0.6 and 0.83 for the half, fundamental, and double cyclotron frequency wave respectively. By comparing these three cases, we observe that ion acceleration by higher-order cyclotron harmonics is generally much less effective than by the wave in the fundamental mode, as it takes a longer time for ions to reach the gyroradius limit in the former cases. The numerically calculated gyroradius limits for integer-harmonic cyclotron waves are consistent with those predicted by \citet{lysak1986}. The limits are determined by the first zeros of $J_{n}(k\rho_{i})$ for cyclotron harmonics. However, no calculation has been reported in the literature for the case of $\omega={\frac{\Omega_i}{2}}$. In Appendix A, we show an analytical derivation of the ion gyroradius limit with the emphasis on this half cyclotron frequency. The analytical predictions are consistent with the numerical results presented. 

Figure~\ref{gyrolimit}d shows the result for waves with two frequencies whose difference is the cyclotron frequency. The ion gyroradius limit still holds for waves with multiple frequencies but with a single wavelength. For Figure~\ref{gyrolimit}e, there are two waves having different $\lambda_{\perp}$, each with half of the original wave amplitude. As a result, the ion gyroradius exceeds the limits set by both perpendicular wavelengths. The ion can be accelerated to a gyroradius more than 2.5 times of the maximum $\lambda_{\perp}$. This is the detrapping effect discussed in \citet{lysak1980}. Note that waves with multiple frequencies that add up to the cyclotron frequency can also accelerate ions (not shown here), which is consistent with previous studies \citep[e.g.][]{temerin1986}.

\begin{figure}[h]
\includegraphics[scale=0.6,width=8cm,trim={1cm 1cm 0 2cm}]{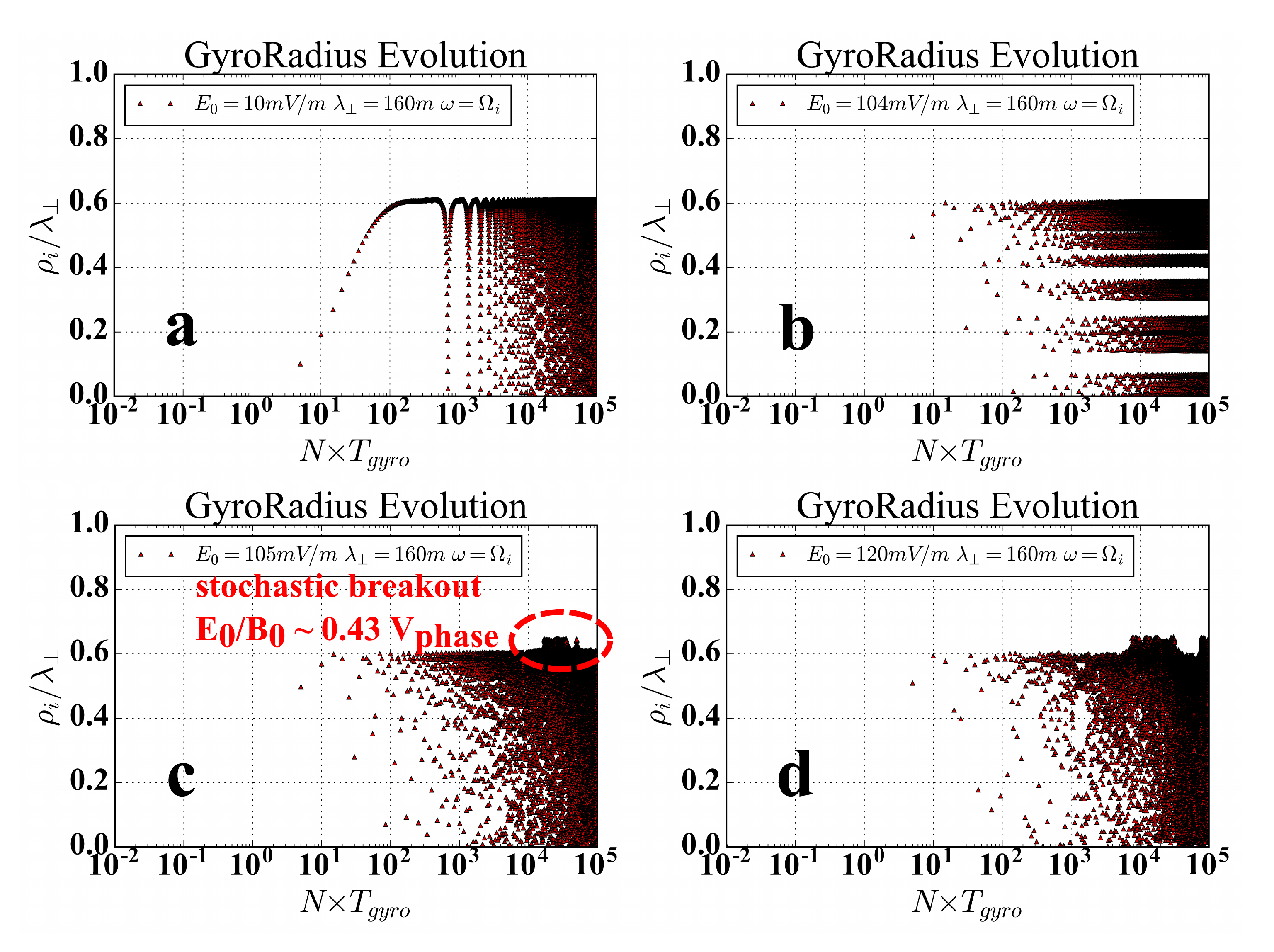}
\caption{Stochastic ``breakout''---ions diffusing beyond the gyroradius limit due to stochastic ion heating---initiates when the EIC wave amplitude increases from 10 mV/m to 120 mV/m, corresponding to $\frac{E_0}{B_0}=0.041V_{\text{phase}}$ to $0.493V_{\text{phase}}$. When ${E_0}$=105 mV/m, as shown in Figure 3c, or equivalently when $\frac{E_0}{B_0}$ approaches $0.43V_{\text{phase}}$ of the wave, the ion gyroradius can be larger than $0.6\lambda_{\perp}$. The magnitude of $B_0$ in the simulation is 40,000 nT. }
\label{breakout}
\end{figure}

One way to break the ion gyroradius limit is to have multiple perpendicular wavelengths in the system as shown in Figure~\ref{gyrolimit}e. The other way, which we found through test particle simulations, is to increase the amplitude of the wave electric field to a degree that ions eventually diffuse out of the perpendicular wavelength limit due to stochastic ion heating. Figure~\ref{breakout} shows how the stochastic ``breakout'' initiates when the ratio of $\frac{E_0}{B_0}$ increases from $0.041V_{\text{phase}}$ ($E_0=$10 mV/m) to $0.493V_{\text{phase}}$ ($E_0$=120 mV/m), where $V_{\text{phase}}$ is the phase speed of the EIC wave. When $\frac{E_0}{B_0}$ approaches $0.43V_{\text{phase}}$ (${E_0}$=105 mV/m) of the wave as depicted in Figure~\ref{breakout}c, ions start to obtain gyroradii larger than $0.6\lambda_{\perp}$. This is a repeatable threshold valid for the EIC wave in the fundamental mode. Such a threshold has not been reported yet. \blue{Although stochastic ``breakout'' only increases the ion energy marginally, resultant detrapping of the ion from the wave potential may contribute to ion loss from a spatially-restricted heating system or a travelling wave. }

\section{e-POP observations}

One of the major objectives of this paper is to investigate essential wave properties, effective perpendicular wavelengths, as well as microscopic heating mechanisms that are responsible for BBELF wave-ion heating as observed from the Enhanced Polar Outflow Probe (e-POP) satellite. The e-POP scientific payload is part of the multipurpose CASSIOPE satellite \citep{yauandjames2015}, launched on 29 September 2013 into a polar elliptical orbit plane with an inclination of $81^\circ$, a perigee of 325 km, and an apogee of 1,500 km. In this section, we show the magnetic field, electric field, and ion observations from the wave-ion heating event on 18 May 2015. Further details on this event can be found in \citet{shen2018}. Relevant instruments onboard e-POP are described separately in \citet{knudsen2015}, \citet{wallis2015}, and \citet{james2015}. The field and particle observations put important constraints on wave electric field amplitudes and ion heating scales, which serve as baselines for the numerical test particle simulations in the next sections. 

\begin{figure}[t]
\includegraphics[scale=0.4,width=8cm,trim={4cm 1cm 0 2cm}]{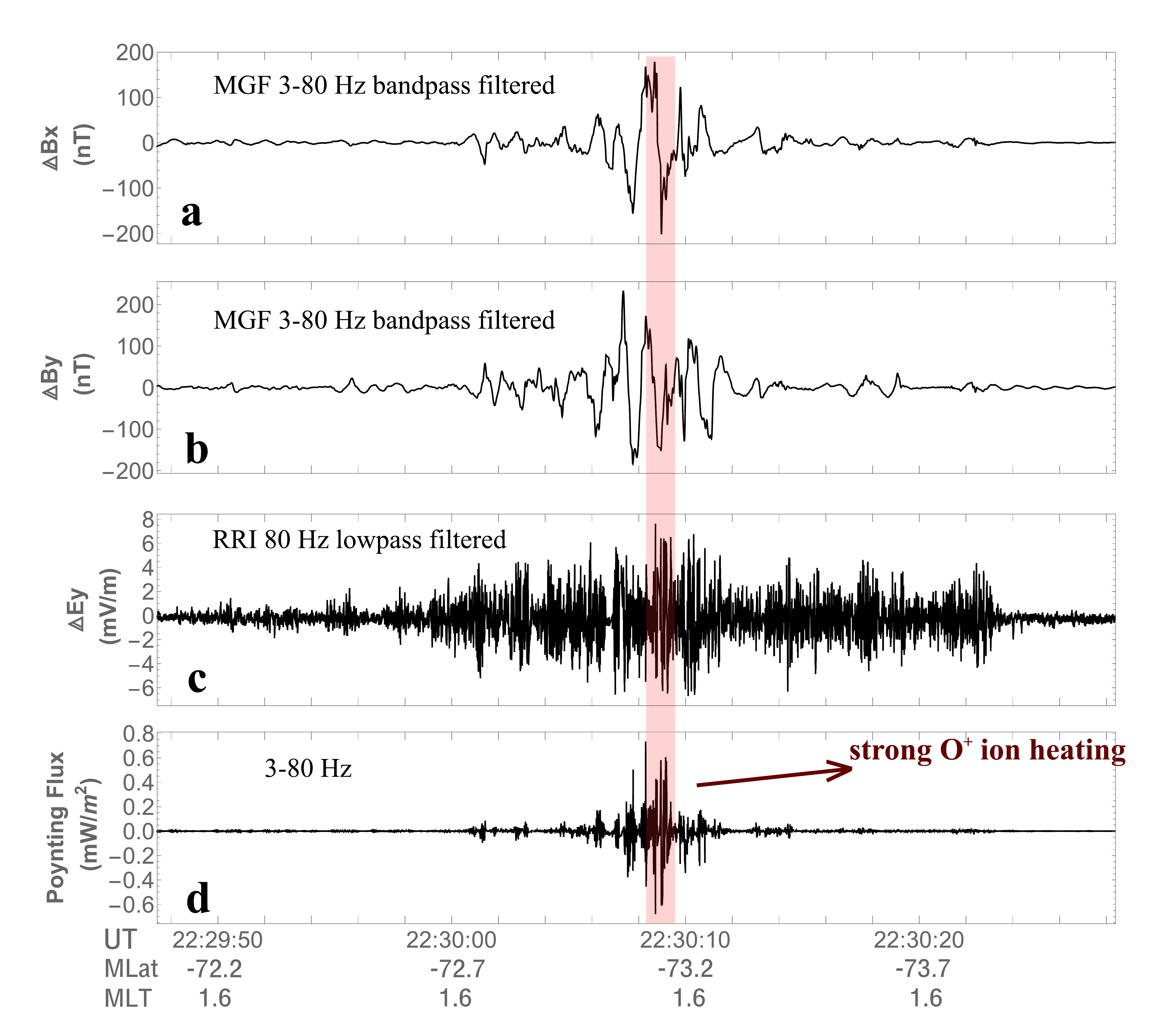}
\caption{Summary of the field observations from e-POP at 410 km altitude between 22:29:47.4 and 22:30:28.4 UT on 18 May 2015. Figure 4a and 4b show magnetic perturbations $B_x$ and $B_y$ in the spacecraft frame. The magnetic field time series are sampled at 160 per second and bandpass filtered to be within the 3-80 Hz range. Figure 4c shows wave electric fields from the Radio Receiver Instrument (RRI) in the cross-track (y) direction. $E_y$ electric fields has been lowpass filtered with a cutoff of 80 Hz. \blue{Time-series electric fields are underestimated at frequencies below 7 Hz due to limitation of the instrument dynamic range.} Note that magnetic fluctuations do not always accompany wave electric fields, meaning the waves are sometimes electrostatic. Figure 4d displays the calculated AC Poynting flux from $B_x$ and $E_y$. Alfv{\'e}nic magnetic fluctuations up to 300 nT, perpendicular electric fields up to 8 mV/m, and Poynting uxes up to 0.8 mW/m$^{2}$ are observed at near 22:30:08 UT. These field fluctuations are colocated with strong O$^+$ ion heating, indicated within the red-shaded region.}
\label{overview}
\end{figure}

Figure~\ref{overview} presents a summary of the field observations from e-POP at 410 km altitude between 22:29:47.4 and 22:30:28.4 UT. Figure~\ref{overview}a and~\ref{overview}b show the magnetic perturbations \blue{$B_x$ and $B_y$} in the spacecraft frame, corresponding to the along-track ($+x$, to the south) and cross-track ($+y$, to the west) component, respectively. Magnetic fields are measured at 160 samples per second (sps) and bandpass-filtered to be within the frequency range of 3-80 Hz. Figure~\ref{overview}c shows wave electric fields measured from the Radio Receiver Instrument (RRI) in the $y$ direction. Electric fields are sampled at 62,500 sps and bandpass-filtered from 7 Hz to 80 Hz in the very low-frequency (VLF) mode. Note that magnetic fluctuations do not always accompany wave electric fields, meaning the waves are sometimes electrostatic. In fact, \citet{shen2018} showed that electric fields within BBELF waves measured by e-POP in this region are mostly linearly polarized perpendicular to the magnetic field with frequencies up to 1 kHz (in their Figure 6). \blue{One clear signature of O$^+$ ion cyclotron waves is present at the exact location of observed ion heating in this case, which will be shown in detail later.} Figure~\ref{overview}d displays the calculated AC Poynting flux from \blue{$B_x$ and $E_y$}. We observe Alfv{\'e}nic magnetic fluctuations up to 300 nT, perpendicular electric fields up to 8 mV/m, and Poynting fluxes up to 0.8 mW/m$^{2}$ at near 22:30:08 UT. These field fluctuations, embedded within BBELF waves, are \blue{also} colocated with strong O$^{+}$ ion heating up to a temperature of 4.3 eV as indicated within the red-shaded region \citep{shen2018}. 

\begin{figure}
\includegraphics[scale=0.5,width=9cm,trim={2cm 1cm 0 2cm}]{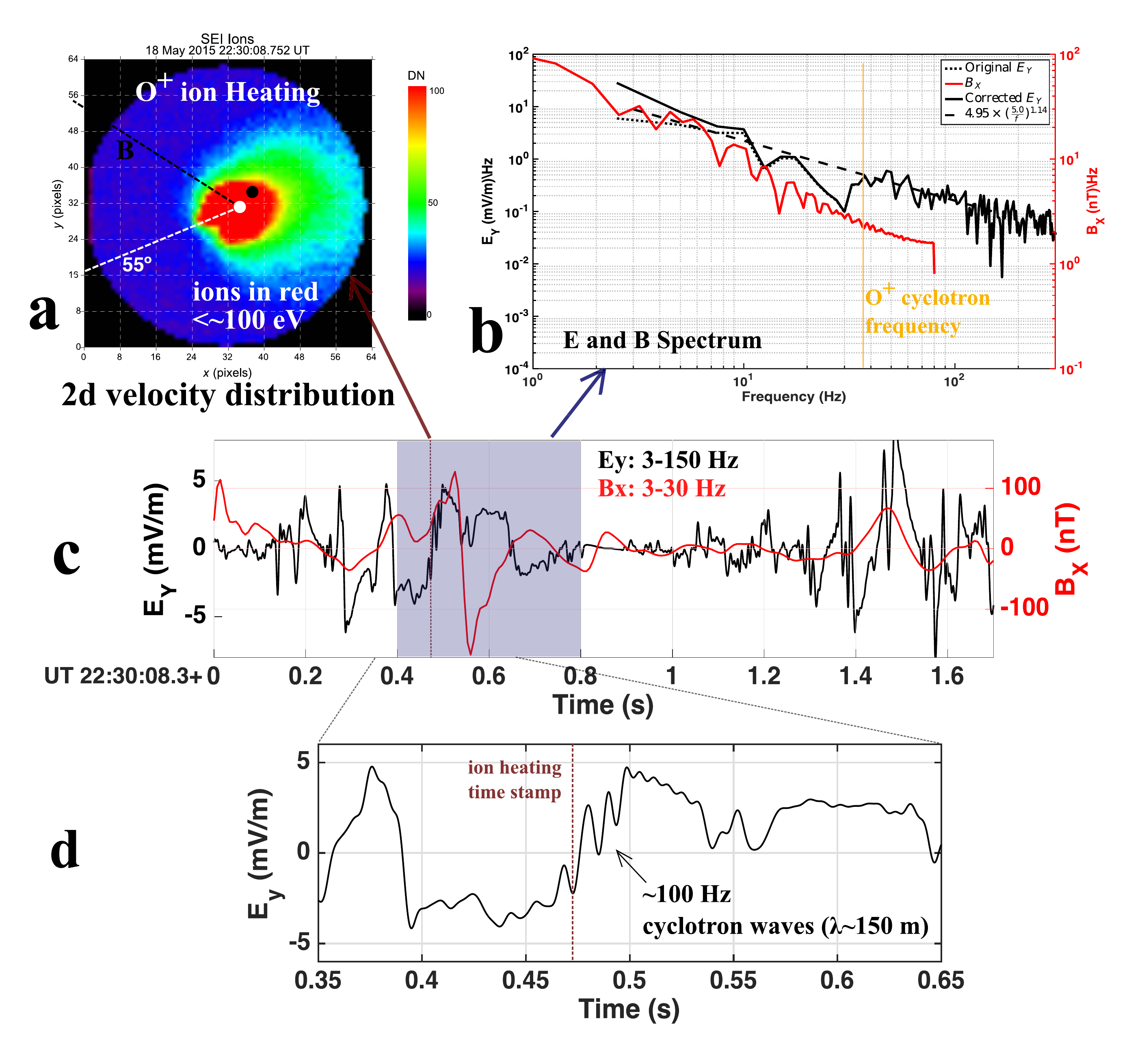}
\caption{Zoomed-in view of the microstructure of ion heating within 1.7 s, along with the BBELF wave spectrum and \blue{short-time-scale electric fields recovered in the frequency range of 3--150 Hz}, for the event shown in Figure~\ref{overview}. Figure 5a shows a heated two-dimensional ion distribution measured from the Suprathermal Electron/Ion Imager (SEI) instrument at 22:30:08.75 UT. This plot is adapted from \citet{shen2018} in their Figure 4c. The projection of the magnetic field direction onto the detector is indicated by the black dashed line. Ion energy increases outward from the centroid of the distribution, which is indicated by the white dot in the image center. The ion energy detectable from SEI goes up to approximately 230 eV at the periphery. Figure 5b shows the \blue{$E_y$ and $B_x$} amplitude-frequency spectra averaged over a 0.4-second measurement period inside an ion heating region. The black solid curve shows the corrected $E_y$. The black dashed curve displays the fitted function form of $E_y=4.95\left({\frac{5}{f}}\right)^{1.14}$. The magnetic perturbation spectrum is shown in the red line. Figure 5c compares the measurements of \blue{$E_y$ and $B_x$} within 1.7 s of ion heating. \blue{The magnetic field fluctuations are bandpass filtered in 3--30 Hz. Note that the time-series $E_y$ in Figure 5c have not been corrected in amplitude as we have done for the spectrum plot in Figure 5b. In Figure 5c, at the time location of O$^+$ ion heating (dark-red dotted grid line), $E_y$ shows fluctuations with a frequency of approximately 100 Hz. These waves are linearly polarized perpendicular to the magnetic field and are consistent with O$^+$ ion cyclotron waves with a perpendicular wavelength of approximately 150 m. A zoomed-in view of these cyclotron waves is shown in Figure 5d. It is worth mentioning that another ion heating event with a temperature of near 2 eV was observed at +1.5 s in Figure 5c, which is not discussed here.}}
\label{microview}
\end{figure}

To demonstrate microstructures of ion heating, BBELF wave spectra, Alfv{\'e}n wave \blue{and cyclotron wave} characteristics, we present a 1.7-second zoomed-in view of the measurements in Figure~\ref{microview}. Figure~\ref{microview}a shows a heated two-dimensional ion enery-angle distribution measured from the Suprathermal Electron/Ion Imager (SEI) instrument at 22:30:08.8 UT. The maximum O$^{+}$ ion temperature, represented by the width of the distribution at $55^\circ$~pitch angle (the white dashed line), is determined to be approximately 4.3 eV, which has been validated through a Monte-Carlo charged particle ray tracing simulation \citep{johnathan2010}. The noteworthy feature in the image is that most of the ion signal lies within the energy of approximately 100 eV, as indicated by the red pixels (not saturated). This is one of the observables we use to compare simulations with observations. In addition, statistical observations from e-POP have shown that ion heating by BBELF waves is associated with ion downflows in the low-altitude (325-730 km) auroral downward current region \citep{shen2018}. \citet{shen2018} applied the``pressure cooker'' ion heating model with down-pointing electric fields in the return current region \citep{gorney1985} to explain the low-altitude e-POP observations. Based on these results, we conclude that O$^+$ ions can be forced to remain within BBELF heating regions long enough for their energies to saturate.. 

Figure~\ref{microview}b presents the \blue{$E_y$ and $B_x$} amplitude-frequency spectra calculated within a 0.4-second measurement period inside an ion heating region. We can correct the magnitude measurement of \blue{$E_y$} below 10 Hz to account for the instrument filter response. This is plotted as the black solid line in Figure~\ref{microview}b. This frequency spectrum up to approximately 120 Hz is fitted with a power-law function $E_y=4.95\left({\frac{5}{f}}\right)^{1.14}$. After taking into account the other component of perpendicular electric field that we do \blue{not} measure, we assume much larger total electric fields as $E=8\left({\frac{5}{f}}\right)^{1.14}$. This is another important baseline setting for simulations. The O$^{+}$ cyclotron frequency is about 38 Hz as indicated by the vertical orange line. \blue{In Figure~\ref{microview}b, the ratios of the corrected $E_y$/$B_x$ in the frequency range of 3--10 Hz suggest that the electromagnetic fluctuations have phase speeds of 400--700 km/s. Assuming O$^{+}$ dominated plasma with an electron density of $10^{11}$ m$^{-3}$ near 400 km altitude, the expected Alfv{\'e}n speed is approximately 700 km/s, which is consistent with the observations.} 

Figure~\ref{microview}c compares the measurements of \blue{$E_y$ and $B_x$} within 1.7 s of the ion heating period. In this case, \blue{the electric field fluctuations are lowpass filtered with a cutoff of 150 Hz.} Although time-series electric field measurements deteriorate at frequencies lower than 10 Hz, we can see in-phase oscillations of electric field and magnetic field perturbations, with an oscillation period of approximately 0.2 \blue{s}, which is identifiable near \blue{+0.5 s and +1.5 s UT} on the time axis. The fairly detailed correlation between \blue{$E_y$ and $B_x$}, \blue{inferred Alfv{\'e}n speeds,} along with the macroscopic Poynting fluxes as shown in Figure~\ref{overview}d, strongly \blue{suggest} that Alfv{\'e}n waves are present within the measured BBELF wave spectrum. \blue{Most interestingly, right at the time location of O$^+$ ion heating (indicated by the dark-red dotted grid line), $E_y$ shows fluctuations with a frequency of approximately 100 Hz. These waves are not observed outside the ion heating time intervals (within 0.2 s centered around +0.5 s UT on the time axis). The waves are linearly polarized perpendicular to the background magnetic field based on two-component electric field ($E_y$ and $E_z$) hodogram analysis (provided in Supporting Information). These features are consistent with electrostatic O$^+$ ion cyclotron waves with amplitudes of up to 5 mV/m and a perpendicular wavelength of approximately 150 m. Such a wave scale is in accord with the gyroradius (180 m) of 100 eV O$^+$ ions. A zoomed-in view of these cyclotron waves is shown in Figure~\ref{microview}d. Therefore, our observations show that both Alfv{\'e}n waves and EIC waves are present within the BBELF wave spectrum and are associated with ion heating observed by e-POP.} 

\blue{Based on previous studies, ion cyclotron waves may be generated by velocity-shear driven instabilities \citep{ganguli1994}, current-driven instabilities \citep{kindel1971}, ion-beam-driven instabilities \citep{hendel1976}, or nonlinear breaking of Alfv{\'e}n waves \citep{seyler1998}. The exact nature of how these kinetic processes produce EIC waves is beyond the scope of this study. }

 \section{Test particle simulation: Ion heating from BBELF waves}
 
Wave frequency measurements from a single spacecraft can be complicated to interpret as to whether they are temporal or spatial structures. This is especially true as the e-POP satellite moves at a speed of 7.8 km/s, which translates to 7.8 Hz for static structures with a perpendicular wavelength of 1 km. In this section, we examine wave-ion heating from BBELF waves, which are either Alfv{\'e}n waves ($\nu=$0.01, different from the earlier simulations to be consistent with common observations in the ionosphere) or EIC waves ($\nu=$1) with varying wavenumber spectra. BBELF waves contain many wave modes, including EIC waves that are slightly above the ion gyrofrequency and ion acoustic waves that are below the ion gyrofrequency \citep{kintner1996,bonnell1996,wahlund1998}. Here we term any electrostatic wave as EIC wave so far as it has a frequency near the ion gyrofrequency. We only consider wave-ion heating in the two-dimensional plane perpendicular to the background magnetic field. Our objective is to find the asymptotic limit of perpendicular ion heating from plasma waves, which justifies our neglecting the effects of parallel transport, including the mirror force and others, that tend to remove hot ions from the heating region. Potential parallel electric fields in Alfv{\'e}n waves, when their perpendicular wavelengths are comparable to the electron inertial length \citep[e.g.][]{goertz1979,lysak1996}, are ignored in our simulations because ion parallel velocities (on the order of 1 km/s) are much less than the Alfv{\'e}n wave parallel phase velocity (on the order of 1000 km/s) so that ions cannot be resonantly accelerated, and their perpendicular temperature is insignificantly affected by parallel electric fields \citep{chaston2004}. 

In the simulation, there are in total 20000 particles, having an initial two-dimensional Maxwellian distribution in velocity space with a temperature of 0.2 eV, and exhibiting a uniformly distribution in the spatial domain $x=[0, 2000]$ m and $y=[-100, 100]$ m. We use the 4th-order Runge-Kutta integrator, as discussed earlier, with a stepsize of $0.01{T_{gyro}}$, to evolve the 20000-particle system numerically. Note that there exists no spatial boundary for an ion's trajectory to evolve in the simulation. The ion temperature is calculated for the whole ion population and is expressed as $T_{\perp}={\frac{m_i}{2{k_B}}}<(v_x-<v_x>)^{2}+(v_y-<v_y>)^{2})>$. Ion temperatures are recorded every 500 steps, corresponding to every $5{T_{gyro}}$. This ensures that we investigate ion dynamics at a constant ion gyro-phase, indicating secular variations. The wave electric field is generally in the form $E_x ={\sum\limits_{40k}^{}{E_0}\text{cos}({k_x}x-{\omega}t+{\phi_{rand}})}$, where 40 different $k$ modes are utilized with random phases ${\phi_{rand}}$. The wave amplitudes are $E_0=8\left({\frac{5}{f}}\right)^{1.14}$. 

In addition, many previous studies suggest that Alfv{\'e}n waves become increasingly non-planar and display two-dimensional (2D) field variations with decreasing scale sizes, especially when $\lambda_{\perp}$ approaches 1 km or less \citep{volwerk1996,chaston2004}. To account for this field topology, we perform additional test runs with both $k_x$ and $k_y$ for Alfv{\'e}n waves, with electric fields specified as $E_x ={\sum\limits_{k_x, k_y}^{}{-k_x \Phi({\omega}t+{\phi_{rand}})\text{cos}({k_x}x)\text{cos}({k_y}y)}}$ and $E_y ={\sum\limits_{k_x, k_y}^{}{k_y \Phi({\omega}t+{\phi_{rand}})\text{sin}({k_x}x)\text{sin}({k_y}y)}}$\citep{chaston2004}, where $k_x$ and $k_y$ are equal to bring out the maximum heating rate and $k\Phi$ is the field amplitude specified according to e-POP observations (the same as $E_0$). \blue{Note that the 2D perpendicular electric field in this equation is curl-free and therefore the ion flow is incompressible.} This is similar to that used in \citet{chaston2004} but excludes variations in the field line ($z$) direction.  
 
 Since observations of ion heating from e-POP are in the ionosphere, we have to take into account ion-neutral collisional effects, which play an important role in limiting ion heating from BBELF waves. Atomic oxygen O$^+$ ions in the ionosphere will primarily experience O$^+$--O resonant charge exchange (RCE) collisions and polarization collisions, where the ion interacts with the electric dipole it induces in the neutral particle. At altitudes of about 400 km, chemical reactions between O$^+$ ions and N$_2$ and O$_2$ neutral gases are of secondary significance. We only include O$^+$--O RCE collisions in the simulations, since long-range polarization collisions only deflect O$^+$ ions that have low relative speeds with neutral particles, and they are less important when $T_{O^+}>$400 K \citep{barakat1983,wilson1994}. Details on how we simulate O$^+$--O collisions are discussed in Appendix B. 
 
 Table~\ref{t1} lists the wave modes, Doppler-shifted frequencies, perpendicular wavelengths, durations of numerical integration, and ion heating features in the eight test runs we performed (see the end of the document). Each test run has been assigned a number indicated in the first column. Broadband emission can be interpreted either as temporal or Doppler-shifted spatial signals in the spacecraft frame, and both contributions will affect the spectrum we observe \citep{stasiewicz2000a, chaston2004}. We assume that much of the broadband frequency spectrum is due to Doppler shift ($\vec{k_{\perp}}{\cdot}\vec{V}$) from a satellite passing through waves with finite $\lambda_{\perp}$ at a relative speed of 7.8 km/s. The only exception is Run-1 (temporal limit $k=0$), where broadband emissions from 0.1$\Omega_i$ to 4$\Omega_i$ are all temporal variations with an infinitely long wavelength. Run-2 (spatial limit $\omega=0$) treats all fluctuations as stationary spatial structures without time oscillations. No heating is observed for these two cases. The movies showing phase space evolution of all the test runs are in Supporting Information (SI). 
 
\begin{figure}
\includegraphics[scale=0.5,width=8cm,trim={4cm 1cm 0 2cm}]{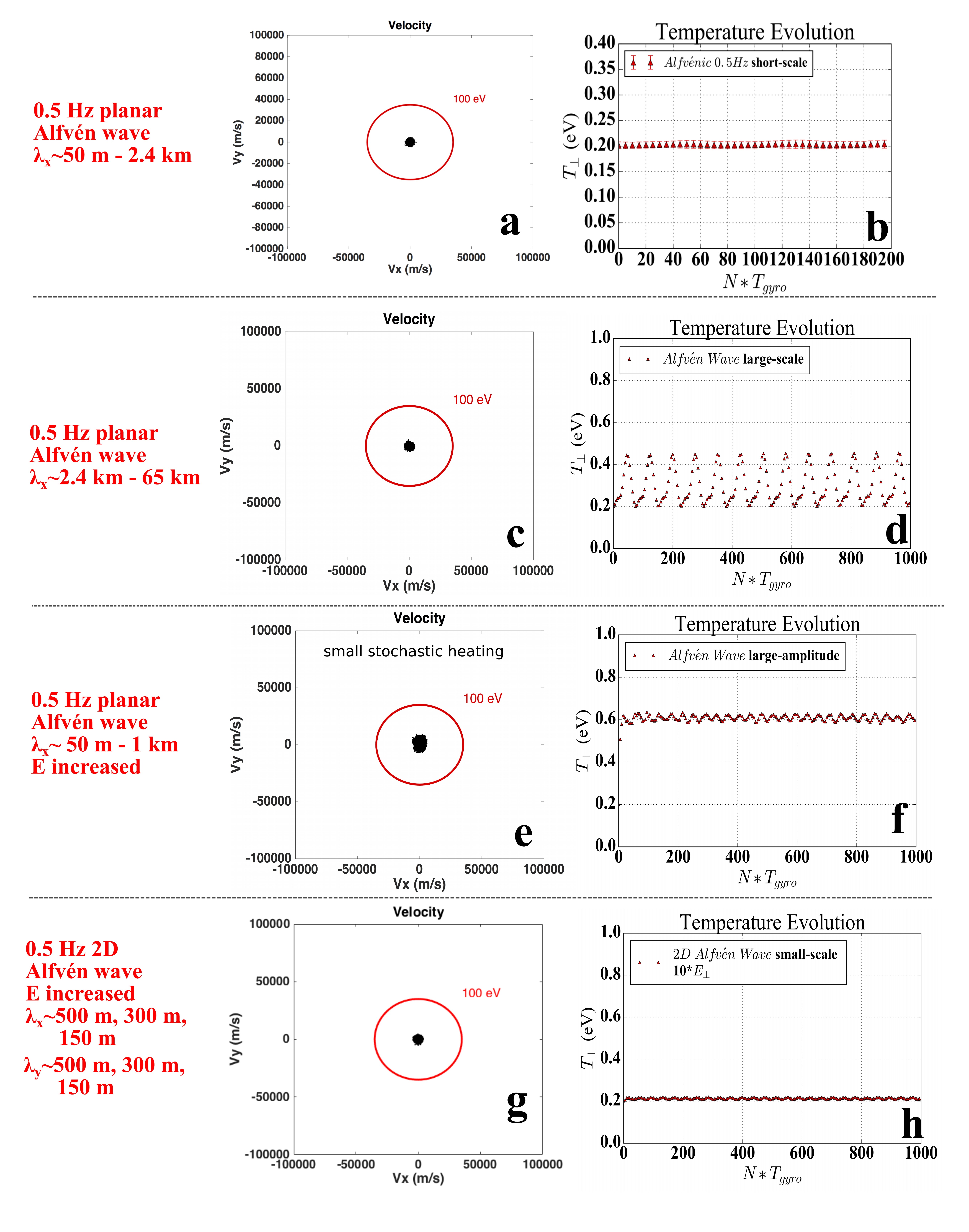}
\caption{Snapshots of ion distribution and temperature evolution for four test runs: (Run-3) 0.5 Hz planar short-scale shear Alfv{\'e}n waves (Figure~\ref{alfven}a and~\ref{alfven}b), (Run-4) 0.5 Hz planar large-scale Alfv{\'e}n waves (Figure~\ref{alfven}c and~\ref{alfven}d); (Run-5) 0.5 Hz planar short-scale Alfv{\'e}n waves with increased amplitudes to initiate stochastic heating (Figure~\ref{alfven}e and~\ref{alfven}f); (Run-8) 0.5 Hz two-dimensional Alfv{\'e}n waves with equal $k_x$ and $k_y$ and with increased amplitudes (Figure~\ref{alfven}g and~\ref{alfven}h), respectively. Snapshots of the velocity distribution are taken at the end of 1000 gyroperiods of integration. Snapshots of temperature are taken every 5 gyroperiods. The red contour (100 eV) in the velocity plots represents the observed O$^+$ ion energy limit from e-POP. }
\label{alfven}
\end{figure}

 Figure~\ref{alfven} shows snapshots of the ion distributions (Figure~\ref{alfven}a,~\ref{alfven}c,~\ref{alfven}e, and~\ref{alfven}g) and temperature evolution (Figure~\ref{alfven}b,~\ref{alfven}d,~\ref{alfven}f, and~\ref{alfven}h) for the test runs of Run-3, Run-4, Run-5, and Run-8. In the case of short-scale shear Alfv{\'e}n waves, we observe no ion acceleration or heating as demonstrated in Figure~\ref{alfven}a and~\ref{alfven}b. On the other hand, large-scale Alfv{\'e}n waves give rise to only a 0.2 eV ion temperature increase through coherent trapping. The periodic oscillation in the ion temperature is due to time oscillation of 0.5 Hz Alfv{\'e}n waves. Short-scale planar shear Alfv{\'e}n waves only generate a small amount of stochastic ion heating up to less than 0.6 eV (Figure~\ref{alfven}e and~\ref{alfven}f) even as we increase the wave amplitudes tenfold. Ion heating is still insignificant, as shown in Figure~\ref{alfven}g and~\ref{alfven}h, when we adopt a different field topology with equal $k_x$ and $k_y$ and with wave amplitudes increased tenfold (approximately 20 mV/m, 10 mV/m, and 5 mV/m for the three wavelengths included).
 
 Our result of 2D Alfv{\'e}n wave ion heating differs from \citet{chaston2004}, who suggested that non-planarity of small-scale Alfv{\'e}n waves can facilitate ion energization up to several keV through coherent trapping at low-altitudes and stochastic ion heating at higher altitudes. \blue{The difference can be explained by the fact that electric field amplitudes of Alfv{\'e}n waves with a perpendicular wavelength of 800 m observed from e-POP (Doppler-shifted to 10 Hz in the satellite frame) are approximately 2 mV/m, which is two orders of magnitude smaller than 200 mV/m electric fields used by \citet{chaston2004} for the same wavelength. It is generally not likely to observe a 200 mV/m electric field at 10 Hz from low-altitude ($<$1000 km) satellites \citep{basu1989,ball1991,wu2020}. As we artificially increase the wave amplitudes of non-planar 2D Alfv{\'e}n waves in our simulation to 200 mV/m (${\lambda_\perp}\sim$500 m), we observe significant ion heating up to 20 eV (not shown here), consistent with the O$^+$ ion temperature obtained by \citet{chaston2004}. The temperature increase will drop to 1 eV with an electric field amplitude of 100 mV/m. Therefore, the relatively weak electric fields at low altitudes limit the role of either planar or 2D Alfv{\'e}n waves in explaining the ion heating we observed through either coherent trapping or stochastic ion heating.}
 
 The insignificant amount of ion heating from small-scale Alfv{\'e}n waves from our simulations is also in contrast to the study of \citet{stasiewicz2000b}, who suggested that stochastic ion heating from small-scale (of the order of 100 m perpendicular wavelength) Alfv{\'e}n waves can explain the bulk ion heating observed at Freja and FAST altitudes (\textgreater1700 km). Similarly, that study used large wave amplitudes of 100 mV/m within tens-of-meter spatial structures to initiate stochastic ion heating. Again, the field amplitudes observed from e-POP and used in the simulations in our paper are an order of magnitude smaller, thus not enough to create significant stochastic ion heating as \citet{stasiewicz2000b} does. \blue{Alfv{\'e}n waves may play an indirect role in ion heating cases at low altitudes, which needs to be investigated in the future.} In addition to the test runs described here, we also run cases that take into account waves with multiple frequencies below the cyclotron frequency. No significant difference is observed compared with the single-frequency Alfv{\'e}nic cases. 
   
\begin{figure}
\includegraphics[scale=0.6,width=8cm,trim={3cm 1cm 0 2cm}]{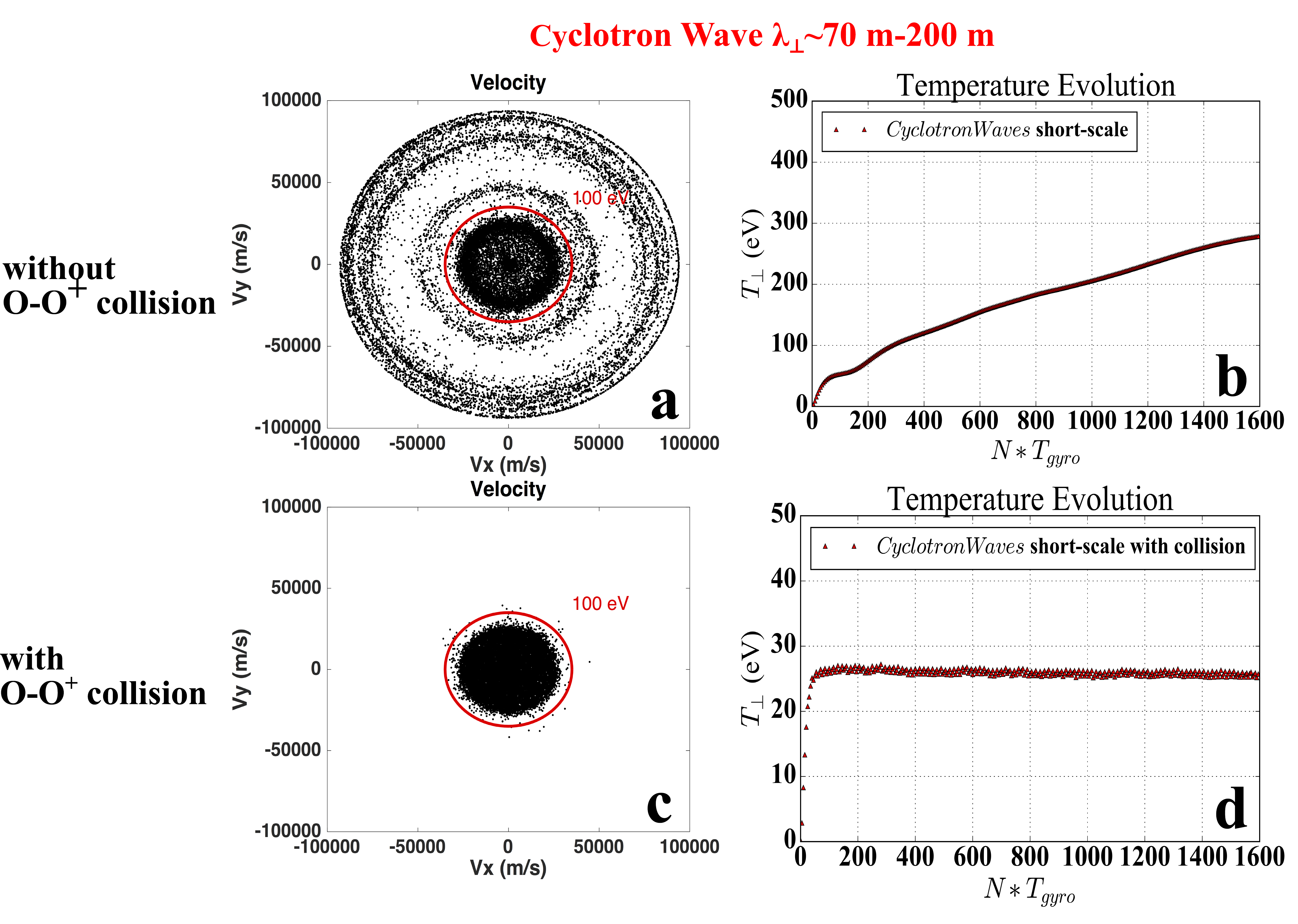}
\caption{Ion velocity distribution and temperature evolution for the test runs (Run-6 and Run-7) of short-scale ($\lambda_{\perp}\sim$70--200 m) cyclotron waves without (Figure~\ref{cyclotron}a and~\ref{cyclotron}b) and with (Figure~\ref{cyclotron}b and~\ref{cyclotron}c) ion-neutral collisions. The format is similar to Figure~\ref{alfven}, except that overall averaged ion temperatures are calculated in the simulation. In Figure~\ref{cyclotron}d, when O$^+$--O collisions collisions are included, ion temperature increases rapidly (within 50 gyroperods, corresponding to near 1 s) to a steady state of 26 eV.}
\label{cyclotron}
\end{figure}

  Figure~\ref{cyclotron} presents results of the other test runs (Run-6 and Run-7) for short-scale ($\lambda_{\perp}\sim$ 70m -- 200 m) cyclotron wave ion heating with (Figure~\ref{cyclotron}a and~\ref{cyclotron}b) and without (Figure~\ref{cyclotron}c and~\ref{cyclotron}d) ion-neutral collisions. In the absence of collisions, small-scale cyclotron waves with wavelengths of less than 200 m effectively heat ions to temperatures of larger than 50 eV within 200 gyroperiods (Figure~\ref{cyclotron}b), corresponding to approximately 5 seconds. Figure~\ref{cyclotron}a illustrates that accelerated ions obtain energies well above the 100 eV observed limit (the red contour line) due to cyclotron heating. However, many ions are trapped at energies less than 100 eV, which can be attributed to the finite wavelength effect. When collisions are present, ions are isotropically restricted within the 100 eV limit and the temperature surges to a steady state of 26 eV within 50 gyroperiods (near 1 second). Figure~\ref{cyclotron}c and~\ref{cyclotron}d demonstrate that the effect of collisions is to limit as well as to heat the system to a steady-state ion temperature. Compared with the other test runs, small-scale cyclotron wave ion heating regulated by ion-neutral collisions can be most effective in heating O$^+$ ions at ionospheric altitudes where e-POP operates. 

\section{Test particle simulation: Cyclotron ion heating regulated by $E^2$, $\lambda_{\perp}$ and collisions}

In this section, we use test particle simulations to explore how ion-neutral collisions, together with $\lambda_{\perp}$ and wave amplitudes, affect ion heating from cyclotron waves in the ionosphere. For multiple test particle simulations, we specify a single cyclotron wave with either an infinite wavelength or a wavelength of 200 m, and with varying amplitudes of 1, 2, 4, or 6 mV/m. Note that the 200 m wavelength is chosen to represent the finite wavelength effect of cyclotron ion heating. We vary the collisional frequency by changing the atomic oxygen O neutral density from 5.7$\times$10$^7$cm$^{-3}$, which is calculated from the MSIS00 model \citep{hedin1991} at 410 km altitude, to half and tenth of this value, calculated at an altitude of 450 km and 540 km respectively. A linearly polarized wave can be decomposed into right-handed and left-handed circularly polarized waves with equal amplitudes and only left-handed polarized waves can interact with left-handedly rotating ions in the magnetic field through cyclotron resonance. Thus, only the left-handed polarized power of cyclotron waves are taken into account for comparison with the model of quasilinear cyclotron heating according to \citet{chang1986}, who suggested that the ion heating rate from cyclotron waves with an infinite wavelength can be expressed as:
\begin{equation}
\label{quasilinear}
\frac{dW}{dt}=\frac{q^2 E^2}{2m_i}
\end{equation}
where $E^2$ is the power spectral density of the left-hand component of electric fields. The cases of cyclotron heating with an infinite wavelength are termed ``temporal'' in the following. 

\begin{figure}
\includegraphics[scale=0.7,width=9cm,trim={3cm 1cm 0 2cm}]{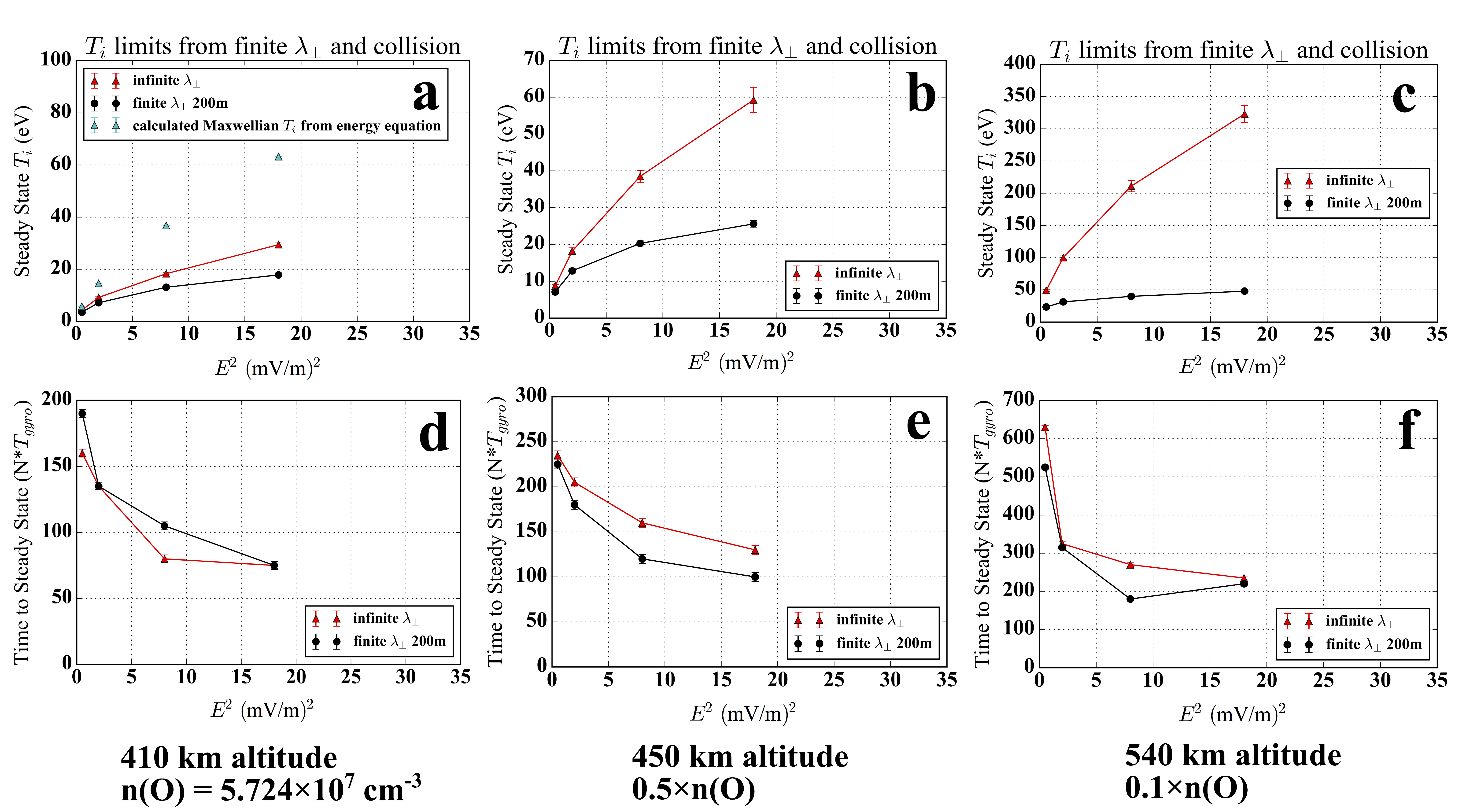}
\caption{Simulation results comparing the magnitude of (Figure~\ref{steadyT}a,~\ref{steadyT}b and~\ref{steadyT}c), and the integration time to (Figure~\ref{steadyT}d,~\ref{steadyT}e and~\ref{steadyT}f), a steady-state ion temperature from cyclotron heating with varying wavelengths, wave powers, and collision frequencies at altitudes from 410 km to 540 km. Steady-state ion temperatures from the cyclotron wave with a perpendicular wavelength of 200 m, and with an infinite wavelength, are indicated by the black dots and lines, and red triangles and lines, respectively. Simulations have been run for 1,000 gyroperiods in all cases. The theoretically calculated steady-state temperatures from Equation~\ref{energy} are displayed as the cyan triangles for left-hand polarized cyclotron waves in the temporal cyclotron heating model.}
\label{steadyT}
\end{figure}

Figure~\ref{steadyT} presents the simulation results comparing the magnitude of, and the integration time to, a steady-state ion temperature from cyclotron heating with varying wavelengths, wave powers, and collisional frequencies at altitudes from 410 km to 540 km. Figure~\ref{steadyT}a,~\ref{steadyT}b and~\ref{steadyT}c show increasing steady-state temperatures, from several eV to tens of eV for $E^2=0.5$ (mV/m)$^2$, with decreasing collision frequencies by tenfold for a 100-km change in altitude. Ion heating limited by the finite wavelength effect (the black dots and lines) has a steady-state temperature smaller than that of temporal cyclotron heating (red triangles and lines) from wave electric fields with the same power. The finite wavelength effect is not discernible when collision frequencies are relatively high and wave powers are relatively weak, but dramatic when collision frequencies are small and wave powers strong, as shown in Figure~\ref{steadyT}a and~\ref{steadyT}c. For temporal cyclotron ion heating, the temperature limits rise by more than ten times, from 4.3 eV to 49 eV when $E^2=0.5$ (mV/m)$^2$, as the collision frequency is lowered by a factor of ten from 410 km to 540 km. In this case, the parameter $\frac{\nu_c}{f_{ci}}$ decreases from about 10$^{-2}$ to 10$^{-3}$, where $\nu_c$ is the collision frequency between neutral O and heated O$^+$ ions in the steady state and $f_{ci}$ is the ion cyclotron frequency in Hz. Overall, collisions play a critical role in regulating ion heating limits in the ionosphere. We observe drastic variations in the ion heating limit with even a 100-km change in altitude for the same wave electric fields. 

The ion heating limit from temporal cyclotron heating with collisions can be theoretically calculated by solving the energy equation: 
\begin{equation}
\label{energy}
\frac{dW}{dt}=\frac{q^2 E^2}{2m_i} - \nu_c \frac{1}{2}m_{i} \bar{v}^2 = 0
\end{equation}
where $\bar{v}$ is the two-dimensional root-mean-square speed in the steady state. Note that $\nu_c$ is constant and can be approximated by $n(O)Q(\bar{v})\bar{v}$ in a steady-state, where $Q(\bar{v})$ is the total collision cross-section. This equation implies that for each effective resonant O$^+$--O collision taking place, the momenta of the two colliding particles exchange completely. The loss of energy due to collisions is balanced by the addition of energy due to ion heating. After we obtain a steady-state root-mean-square speed $\bar{v}$ based on Equation~\ref{energy}, we derive the steady-state ion temperature of a two-dimensional Maxwellian distribution based on the assumed relation:
 \begin{equation}
\label{vtoT}
\bar{v}=\sqrt{\frac{2kT_{i}}{ m_i}}
\end{equation}
We display the theoretically calculated steady-state temperatures for cyclotron waves with different powers as the cyan triangles in Figure~\ref{steadyT}a. The simulated (red) and theoretically calculated (cyan) temporal cyclotron heating limits agree well when $E^2$ is below 2 (mV/m)$^2$. When the wave power becomes larger, the calculated Maxwellian temperatures become much greater than, and deviate from, the simulated limits. This is because the simulated ion distribution becomes increasingly non-Maxwellian and more flat-shaped due to larger electric fields, rendering Equation~\ref{vtoT} incorrect.

Figure~\ref{steadyT}d,~\ref{steadyT}e and~\ref{steadyT}f display integration times to reach a steady-state temperature as the collision frequency declines. As $E^2$ and the collision frequency increase, ions reach a steady-state temperature more rapidly, generally in tens of to hundreds of cyclotron periods, corresponding to approximately 1 s to 10 s. Finite wavelength effects heat ions to a steady state temperature more efficiently when collision frequencies are small (Figure~\ref{steadyT}e and ~\ref{steadyT}f). In Figure~\ref{steadyT}d, temporal cyclotron heating accelerates ions more effectively than with a wave having a finite wavelength, and collisions prevail over finite wavelengths in limiting the system to a steady state. Comparing Figure~\ref{steadyT}d with Figure~\ref{steadyT}e, as the collision frequency decreases by half, we observe a cross-over between collisional and finite wavelength effects in heating ions to a steady-state temperature. In general, the effects of $\lambda_{\perp}$, wave power and collision collectively determine the magnitude and efficiency (time to reach a steady-state temperature) of ion heating in the ionosphere. 

Note that our neglecting of the magnetic mirror force may overestimate the ion heating limits only slightly, since within the period of 10 s, cold ionospheric ions cannot travel far along the field line and transfer of energy from perpendicular to parallel via the mirror force is quite limited. \blue{For example, ionospheric O$^+$ ions with a parallel temperature of 1 eV have parallel speeds of the order of 3 km/s. On a time scale of 10 s, an O$^+$ ion flowing upward due to the magnetic mirror force will travel a vertical distance of 30 km. The perpendicular energy of the ion with an initial value of 10 eV will decrease by only 0.1 eV.} In addition, in light of evidence of a low-altitude ``pressure cooker'' effect \citep{shen2018}, ions may well be trapped in the heating region long enough to reach the heating limits derived here under the assumption of negligible field-aligned transport.


\section{Summary and Conclusions}

In this paper, we have examined the physics of ion heating from Alfv{\'e}n waves ($\omega$$\ll$$\Omega_i$) and cyclotron waves ($\omega=\Omega_i$) using test particle simulations, with an objective to explain how BBELF waves heat ionospheric ions as observed from the e-POP satellite in the low-altitude ($\sim$400 km) ionosphere. We have investigated the effects of finite perpendicular wavelengths, wave amplitudes, and ion neutral collisions, on ionospheric ion heating. The results of our test particle simulations in this paper show the following: 
\begin{enumerate}
\item The numerical onset thresholds of stochastic ion heating from the EIC wave and Alfv{\'e}n wave, as illustrated by the Poincar{\'e} surface of section plots, are consistent with previous studies \citep{karney1978,lysak1986,bailey1995, chaston2004}. One important characteristic of stochastic ion heating is that ions undergoing stochastic acceleration can still be limited by the large potential structure of the wave, or similarly by the perpendicular wavelength.
\item The ion gyroradius limit for the EIC wave with half of the cyclotron frequency is 0.28$\lambda_{\perp}$, which has been both numerically and analytically derived. The ion gyroradius limit from a single EIC wave can be surpassed either through adding waves with different $\lambda_{\perp}$, or through stochastic ``breakout'', when the wave electric field amplitude satisfies $\frac{E_0}{B_0}\geq0.43V_{\text{phase}}$, where $V_{\text{phase}}$ is the wave phase velocity.
\item In contrast to previous studies focusing on higher altitudes \citep{stasiewicz2000b,chaston2004}, our simulations indicate that both planar and non-planar small-scale (\textless1 km) Alfv{\'e}n waves cannot explain ion heating in our case. In order to reproduce ion distributions observed at e-POP altitudes, the most effective ion heating mechanism from BBELF waves is through collisional cyclotron heating by short-scale EIC waves with $\lambda_{\perp}\leq200$ m.  
\item The interplay between finite perpendicular wavelengths, wave amplitudes, and ion-neutral collision frequencies collectively determine the ionospheric ion heating limit, which begins to decrease sharply with decreasing altitude below approximately 500 km altitude, where the ratio $\frac{\nu_c}{f_{ci}}$ becomes significant ($>$10$^{-3}$) with decreasing altitude. 
\end{enumerate}
Our model of ion heating by Alfv{\'e}n waves is limited to the perpendicular plane. Ion dynamics in the direction parallel to the magnetic field is not included; as a result, the magnetic mirror force and parallel electric fields are not included. While effects of parallel dynamics on ion heating are not included, they are expected to be small in the low-altitude (410 km) ionosphere as discussed above. 

\newpage
\appendix

\section{Ion gyroradius limit from a single EIC wave}

\citet{lysak1986} has given the derivation of the ion gyroradius limit for EIC waves with ${\omega}={n\omega_{c}}$. Here we present an analytical derivation of the ion gyroradius limit using the gyro-orbit-averaging method for EIC waves with half of the cyclotron frequency, since this has not been reported elsewhere in the literature. 

Following the main text, we assume a single coherent EIC wave with a form $E={{E_0}{cos({k_{\perp}}x-{\frac{\Omega_i}{2}}t)}}$ in a background magnetic field $\vec{B}$ in the $\hat{z}$ direction. Since the wave is electrostatic in nature, the wave scalar potential may be expressed as $\Phi={{-\frac{E_0}{k_{\perp}}}\text{sin}({k_{\perp}}x-{\frac{\Omega_i}{2}}t)}$. The ion can be viewed as being trapped by the wave potential, therefore the total energy of the wave-ion system is conserved. The maximum of the ion's kinetic energy corresponds to the minimum of the wave potential energy, which translates to finding zeros of the wave electric field. Considering secular increase of the ion's gyroradius, we shall obtain a wave electric field that is averaged over one ion gyro-orbit. The electric field is first rewritten in the reference frame of the guiding center: 
\begin{equation}
\label{eq:electricfield}
E = E_0{\text{cos}({k_{\perp}}{x_0}+{k_{\perp}}{\rho_i}\text{cos}({\Omega_i}t)-{\frac{\Omega_i}{2}}t)}
\end{equation}
where $x_0$ is guiding center coordinate, $\rho_i$ is the ion gyroradius. From trigonometric identities, this can be expressed as: 
\postdisplaypenalty=0
\begin{multline}
\label{eq:expand1}
E = E_0\{{\text{cos}({k_{\perp}}{x_0})\text{cos}({\frac{\Omega_i}{2}}t)\text{cos}({k_{\perp}}{\rho_i}\text{cos}({\Omega_i}t))} \\
+{\text{cos}({k_{\perp}}{x_0})\text{sin}({\frac{\Omega_i}{2}}t)\text{sin}({k_{\perp}}{\rho_i}\text{cos}({\Omega_i}t))}\\
-{\text{sin}({k_{\perp}}{x_0})\text{cos}({\frac{\Omega_i}{2}}t)\text{sin}({k_{\perp}}{\rho_i}\text{cos}({\Omega_i}t))} \\
+ {\text{sin}({k_{\perp}}{x_0})\text{sin}({\frac{\Omega_i}{2}}t)\text{cos}({k_{\perp}}{\rho_i}\text{cos}({\Omega_i}t))}\}
\end{multline}
This form can be further expanded using the Bessel function identities equations ${\text{cos}(z{\text{cos}\theta})}={J_0(z)+2{\sum\limits_{n=1}^{\infty}(-1)^{n}J_{2n}(z)\text{cos}(2n\theta)}}$, ${\text{sin}(z{\text{cos}\theta})}=-2{\sum\limits_{n=1}^{\infty}(-1)^{n}J_{2n-1}(z)\text{cos}((2n-1)\theta)}$. After inserting these Bessel identities into the equation and integrate the electric field within one ion gyro-orbit $<E> = {\frac{1}{2\pi}}{\int^{2\pi}_0}E\ d({\Omega_i}t)$, we have the wave electric field in the final form:
\begin{multline}
\label{eq:expand2}
<E> = E_0{\text{cos}({k_{\perp}}{x_0})}\{-{\frac{4}{3\pi}}{J_1}({k_{\perp}}{\rho_i})\\
+{\frac{4}{35\pi}}{J_3}({k_{\perp}}{\rho_i})-{\frac{4}{99\pi}}{J_5}({k_{\perp}}{\rho_i})+\cdots\}\\
+E_0{\text{sin}({k_{\perp}}{x_0})}{\{\frac{2}{\pi}}{J_0}({k_{\perp}}{\rho_i})+{\frac{4}{15\pi}}{J_2}({k_{\perp}}{\rho_i})\\
-{\frac{4}{63\pi}}{J_4}({k_{\perp}}{\rho_i})+\cdots\}
\end{multline}
The maximum magnitude of the averaged electric field is found when setting $\text{sin}({k_{\perp}}{x_0})=\text{cos}({k_{\perp}}{x_0})=\frac{\sqrt{2}}{2}$, in which case the wave potential magnitude is at its largest. The maximum ion kinetic energy, or equivalently the ion gyroradius limit, is obtained from the first zero of $<E>$. Since the terms after ${J_5}({k_{\perp}}{\rho_i})$ are negligibly small, we can truncate the series and find the solution to be ${k_{\perp}}{\rho_i}=1.8$, which translates to the ratio $\frac{\rho_i}{\lambda_{\perp}}=0.28$. This analytical solution is consistent with the numerical result shown in Figure~\ref{gyrolimit}a.

Using the same procedure and change the wave frequency to $\omega={\Omega_i}$, $2{\Omega_i}$, $3{\Omega_i}$ and $4{\Omega_i}$, the resulting averaged wave electric fields are, respectively, $-E_0{\text{sin}({k_{\perp}}{x_0})}J_1({k_{\perp}}{\rho_i})$, $-E_0{\text{cos}({k_{\perp}}{x_0})}J_2({k_{\perp}}{\rho_i})$, $E_0{\text{sin}({k_{\perp}}{x_0})}J_3({k_{\perp}}{\rho_i})$, and $E_0{\text{cos}({k_{\perp}}{x_0})}J_4({k_{\perp}}{\rho_i})$. The first zeros of $J_n({k_{\perp}}{\rho_i})$ correspond to $\frac{\rho_i}{\lambda_{\perp}}=0.61,\ 0.82,\ 1.0,\ 1.2$ for $n=1,\ 2,\ 3\ ,4$. This result is in agreement with \citet{lysak1986}. 

\section{Simulating O--O$^+$ collisions}
We use an approximate method of O$^+$--O collisions, which is essentially that used by \citet{barakat1983} and \citet{wilson1994}. The probability that a particle suffers no collision in the time interval $t$ overall follows Poisson statistics $P_{nc}(t)=e^{-\nu_c t}$. The probability of collision is therefore $P_{c}(t)=1-e^{-\nu_c t}$. $\nu_c$ depends on the neutral O density $n(O)$, RCE collisional cross section, and the relative speed $g$ between O$^+$ and O. Ideally, the overall probability of collision during the time $dt$ can be written in a general form as $P_{c}(t)=1-e^{-dt n(O) \int_{}^{} {\sigma (g) g f(g) dg}}$, where $\sigma (g)$ is the differential cross-section of RCE collisions and $f(g)$ is normalized so that $\int_{}^{} f(g) dg = 1$. We adopt the same approximate method as that used by \citet{wilson1994}, where the integral in the exponential is replaced by $n(O)Q(g)g$ ($Q(g)$ is the total cross section) and $g$ is found at each time step. We assume the background neutral O gas is near stationary so that for each RCE collision the O$^+$ velocity is set to 0 with a probability of 0.5. The neutral O density is obtained from MSIS00 model \citep{hedin1991} using geophysical parameters relevant to the ion heating event reported by \citet{shen2018}. If $\nu_c$ is a constant, the relationship between successive collisional time interval $dt$ and $P_{c}(t)$ can be simplified as $dt=-\frac{1}{n(O)Q(g)g}\ln(1-P_{c}(dt))$. To take into account speed dependence of the collision time $dt$, we use the ``null collision'' method \citep{lin1977,winkler1992}. The simulation steps are:
\begin{enumerate}
\item Specify a very small constant $dt$, which represents the maximum $\nu_c$ possible to occur. We use a $dt$ of about 0.03 s in this paper compared with 0.05 s used by \citet{wilson1994}. 
\item Assume collisions occur at this constant $dt$. Calculate the impact parameter $b$ based on $b=\sqrt{\frac{-\ln(1-r)}{\pi n(O) g dt}}$, where $r$ is a psudorandom number between 0 and 1 representing the collision probability within time $dt$. 
\item Calculate a critical impact parameter $b_{cr}$, which is determined by the experimental $Q(g)$ measured for RCE collisions. We use the same phenomenological RCE model as in \citet{wilson1994}. $b_{cr}$ is calculated for each $g$ such that the total cross section for RCE satisfies $Q(g)= (10.995-0.95\log_{10}g)^2 \times 10^{-16}$ cm$^2$. The probability of charge exchange to occur during each collision is 0.5. 
\item Compare $b$ with $b_{cr}$. If $b \leq b_{cr}$, a real charge exchange collision occurs during this time interval $dt$ and the O$^+$ ion velocity is set to 0; if $b>b_{cr}$, the assumed collision is ``null'' and the O$^+$ ion velocity is not changed.
\item Proceed to the next $dt$ and repeat the above process.
\end{enumerate}
Using this approach, we find that each test particle within a heated (4 eV for example) ion population experiences approximately 10 collisions within the integration time of 1000 gyroperiods. This magnitude of O$^+$--O collision frequency is comparable with values estimated by \citet{schunk_nagy2009} assuming a reduced temperature of approximately 2 eV.
\newpage
\begin{acknowledgments}
This work was supported by an Eyes High Doctoral Recruitment Scholarship from the University of Calgary, the Natural Sciences and Engineering Research Council of Canada (NSERC), and the Canadian Space Agency (CSA). e-POP was funded with support from CSA and MDA Corporation. In 2017 e-POP has been incorporated as one part (Swarm Echo) of the Swarm constellation funded by the European Space Agency. e-POP data are accessible through http://epop-data.phys.ucalgary.ca/. \blue{Simulation data and codes necessary to reproduce the results are available from PRISM Dataverse at University of Calgary's Data Repository through https://doi.org/10.5683/SP2/PWYYZQ}.
\end{acknowledgments}

\end{article}


%
%
%
%
%
%

\begin{sidewaystable}
\centering
\caption {Parameters for the 8 test particle simulation runs.}
\label{t1}
\begin{tabular}{|c|l|c|p{2cm}|p{2cm}|p{2.2cm}|p{4.2cm}|}
\hline
\textbf{Number} & \textbf{Wave Mode} &  \textbf{$\omega_{Doppler-shifted}$}  & \textbf{$\lambda_{\perp}$}  & \textbf{Initial ${k_{\perp}}{\rho_i}$} &  \textbf{Integration Time} & \textbf{Heating feature} \\
\hline
\textbf{Run-1} &$Temporal\ Limit$ $k=0$  & 0.1$\Omega_i$-4$\Omega_i$ & $\infty$ m & 0 &$200\ T_{gyro}$ & no heating\\

\textbf{Run-2} &$Spatial\ Limit$ $\omega=0$ & 0.1$\Omega_i$-4$\Omega_i$ & 50 m -  2000 m & 0.02-0.8 &$200\ T_{gyro} $& no heating\\

\textbf{Run-3} &1D $Alfv{\acute{e}}n$ 0.5 Hz & 0.1$\Omega_i$-4$\Omega_i$ & 50 m - 2400 m & 0.017-0.8 &$200\ T_{gyro}$& no heating \\

\textbf{Run-4} &1D $Alfv{\acute{e}}n$ 0.5 Hz & 0.01$\Omega_i$-0.1$\Omega_i$ & 2400 m - 65600 m& 0.0006-0.017 & $1000\ T_{gyro}$ &  trapping \\

\textbf{Run-5} &1D $Alfv{\acute{e}}n$ 0.5 Hz $Stochastic$ & 0.2$\Omega_i$-4$\Omega_i$  & 50 m - 1000 m & 0.04-0.8 & $1000\ T_{gyro}$& small stochastic heating  \\

\textbf{Run-6} &$Cyclotron$ $\Omega_i$& 0.1$\Omega_i$-2.0$\Omega_i$ & 70 m - 200 m & 0.22-0.6 & $1600\ T_{gyro}$& strong cyclotron heating\\ 

\textbf{Run-7} &$Cyclotron$ $\Omega_i$+Collision& 0.1$\Omega_i$-2.0$\Omega_i$ & 70 m - 200 m & 0.22-0.6 & $1600\ T_{gyro}$& limited cyclotron heating\\ 

\textbf{Run-8} &2D $Alfv{\acute{e}}n$ with 10*$E_{\perp}$ & 0.42$\Omega_i$,0.69$\Omega_i$,1.4$\Omega_i$  & 500 m, 300 m, 150 m & 0.04-0.8 & $1000\ T_{gyro}$& no heating  \\
\hline
\end{tabular}
\end{sidewaystable}
\clearpage

\end{document}